\documentclass[final,3p,times,twocolumn]{elsarticle}
\pdfoutput=1 
\DeclareGraphicsExtensions{.pdf}

\usepackage{graphicx}
\usepackage{atlasphysics}

\usepackage{hyperref}
\usepackage{dcolumn}
\usepackage{multirow}
\usepackage{epstopdf}
\usepackage{lineno}
\usepackage{xspace}
\usepackage{longtable}
\usepackage{amsmath}

\newcommand{\atlasnote}[1]{\def\@atlasnote{#1}}

\biboptions{numbers,square,comma,sort&compress}
\usepackage{hypernat}

\newcommand{\hmm}{\ensuremath{H \rightarrow \mu^+\mu^-}\xspace}
\newcommand{\lumi}{\ensuremath{24.8~\rm{fb}^{-1}}\xspace}
\newcommand{\lumiEa}{\ensuremath{20.3 \pm 0.6~\rm{fb}^{-1}}\xspace}
\newcommand{\lumiEb}{\ensuremath{4.5 \pm 0.1~\rm{fb}^{-1}}\xspace}

\newcommand{\zgam}{\ensuremath{Z/\gamma^{*}}\xspace}
\newcommand{\ptmm}{\ensuremath{p_{\rm T}^{\mu^{+}\mu^{-}}}\xspace}

\newcommand{\hrange}{\ensuremath{120\mbox{--}150~\rm{GeV}}\xspace}
\newcommand{\frange}{\ensuremath{110\mbox{--}160~\rm{GeV}}\xspace}

\usepackage{preprintcover}  
\PreprintCoverPaperTitle{Search for the Standard Model Higgs boson decay to $\mu^{+}\mu^{-}$ with the ATLAS detector}
\PreprintIdNumber{CERN-PH-EP-2014-131}
\PreprintCoverAbstract{A search is reported for Higgs boson decay to $\mu^{+}\mu^{-}$ using data with an integrated luminosity of \lumi collected with the ATLAS detector in $pp$ collisions at $\sqrt{s}=$7~and~8~TeV at the CERN Large Hadron Collider. The observed dimuon invariant mass distribution is consistent with the Standard Model background-only hypothesis in the 120--150~GeV search range. For a Higgs boson with a mass of 125.5 GeV, the observed (expected) upper limit at the 95\% confidence level is 7.0 (7.2) times the Standard Model expectation. This corresponds to an upper limit on the branching ratio BR(\hmm) of $1.5\times10^{-3}$.}  
\PreprintJournalName{Physics Letters B}  

\clearpage

\begin{document}
\setlongtables

\title{Search for the Standard Model Higgs boson decay to $\mu^{+}\mu^{-}$ with the ATLAS detector}

\begin{abstract}
A search is reported for Higgs boson decay to $\mu^{+}\mu^{-}$ using data with an integrated luminosity of \lumi collected with the ATLAS detector in $pp$ collisions at $\sqrt{s}=$7~and~8~TeV at the CERN Large Hadron Collider. The observed dimuon invariant mass distribution is consistent with the Standard Model background-only hypothesis in the 120--150~GeV search range. For a Higgs boson with a mass of 125.5 GeV, the observed (expected) upper limit at the 95\% confidence level is 7.0 (7.2) times the Standard Model expectation. This corresponds to an upper limit on the branching ratio BR(\hmm) of $1.5\times10^{-3}$.
\end{abstract}
\maketitle

\section{Introduction}

The Standard Model (SM) describes a wide range of particle physics phenomena to a high degree of precision. In the SM, the Brout--Englert--Higgs (BEH) mechanism~\cite{Englert:1964et,Higgs:1964pj,Guralnik:1964eu} spontaneously breaks the electroweak (EW) gauge symmetry and generates masses for the $W$ and $Z$ gauge bosons as well as for the charged fermions via Yukawa couplings~\cite{Weinberg:1967tq,Glashow:1961tr,Salam:1968rm}. In searches for the Higgs boson predicted by the BEH mechanism, the ATLAS and CMS collaborations have discovered a new particle, via decays into gauge bosons~\cite{ATLAS:2012gk,CMS:2012gu}, with a mass of approximately 125.5~GeV and measured properties consistent with those predicted by the SM~\cite{ATLAS:2013wqa,ATLAS:2013xqa,CMS:2014vua,CMS:2013lba}.

Higgs boson decays to $b\bar{b}$, $\tau^+\tau^-$ and $\mu^{+}\mu^{-}$ can be measured at the LHC with their SM branching ratios proportional to the squares of the fermion masses. The SM branching ratio for the $\hmm$ decay is $21.9 \times 10^{-5}$ for a Higgs boson mass ($m_{H}$) of 125 GeV~\cite{LHCHiggsCrossSectionWorkingGroup:2011ti,LHCHiggsCrossSectionWorkingGroup:2012vm}.  The $\hmm$ decay has a clean final state signature that allows a measurement of the Higgs boson coupling to second-generation fermions. The dominant irreducible background is the $\zgam \rightarrow \mu^+\mu^-$ process, which has an approximately three orders of magnitude higher production rate compared to that of the expected signal. 

In this Letter a search for the $\hmm$ decay of the SM Higgs boson is presented.
This search for the presence of a narrow $\hmm$ resonance,
with a signal width determined by experimental resolution, is performed
by fitting the invariant mass distribution in the \frange 
region. This range allows determining background
shape and normalisation and setting a limit on the dimuon decay
of the SM Higgs boson with a mass of 125.5 GeV.
Section~\ref{sec:data} gives a description of the experimental setup and summarises the data sample and Monte Carlo (MC) simulation samples used to model the signal process and to develop an analytical model for the background processes. Sections~\ref{sec:event} and~\ref{sec:categorisation} describe the event selections and categorisation. Analytical models used to describe invariant mass distributions for signal and background processes are discussed in Section~\ref{sec:signalbg}, and systematic uncertainties are detailed in Section~\ref{sec:systematics}. The results are presented in Section~\ref{sec:results}.

\section{Experimental setup, data and simulated samples}\label{sec:data}

This search is performed on the data sample recorded in 2011 and 2012 by the ATLAS detector in $pp$ collisions  at the LHC at $\sqrt{s}=$7~and~8~TeV, respectively. ATLAS~\cite{atlas-det} is a general-purpose particle detector with a cylindrical geometry
 and consists of several subdetectors surrounding the interaction point and covering almost the full solid angle.\footnote{ATLAS uses a right-handed coordinate
system with its origin at the nominal interaction point (IP) in the
centre of the detector and the $z$-axis along the beam pipe. The
$x$-axis points from the IP to the centre of the LHC ring, and the
$y$-axis points upward. Cylindrical coordinates $(r,\phi)$ are used
in the transverse plane, $\phi$ being the azimuthal angle around the
beam pipe. The pseudorapidity is defined in terms of the polar angle
$\theta$ as $\eta=-\ln\tan(\theta/2)$.}
The trajectories and momenta of charged particles 
are measured within the pseudorapidity range of $|\eta|<$~2.5 by multi-layer silicon pixel and microstrip detectors as well as a transition radiation tracker. 
The tracking system is immersed in a 2~T magnetic field produced by a superconducting solenoid, and is surrounded by a high-granularity liquid-argon (LAr) 
electromagnetic sampling calorimeter covering $|\eta|<3.2$. An iron and scintillator tile hadronic 
calorimeter provides coverage in the range $|\eta|<1.7$. The LAr calorimeters 
also provide  measurements of electromagnetic and hadronic energy deposits in the region $1.5<|\eta|<4.9$.
The muon spectrometer surrounds the calorimeters and consists of three large air-core superconducting magnets with a torodial field of 0.5~T in the barrel region and 1~T in the forward regions, and a system of precision tracking chambers; it provides coverage for $|\eta|<2.7$. The muon spectrometer also includes fast detectors for triggering covering $|\eta|<2.4$.
 
Events were recorded with a trigger~\cite{Aad:2012xs} requiring at least one muon candidate with transverse momentum $\pt > 18$~GeV for 7~TeV data and $\pt > 24$~GeV for 8~TeV data. This trigger is approximately 70\% efficient for muons with $|\eta| < 1.05$ and approximately 90\% efficient for muons with $|\eta| > 1.05$. The differences in the efficiency are mostly due to the different geometrical acceptance of the muon trigger system in these regions. 
For 8~TeV data, events were also recorded with a dimuon trigger requiring at least two muon candidates with transverse momenta of $\pt > 18$~GeV and $\pt > 8$~GeV. 
After applying data quality requirements, the total integrated luminosity of the selected data sample is \lumiEb for 7~TeV data and \lumiEa for 8~TeV data. 
The associated systematic uncertainties are summarised in Section~\ref{sec:systematics}.

At the LHC, SM Higgs boson production is dominated by the gluon fusion (\text{ggF}) process with the next two most important contributions arising from the vector boson fusion (\text{VBF}) process and production in association with vector bosons (\text{VH}). The signal MC samples are generated in 5~GeV steps of the Higgs boson mass from 120~GeV to 150~GeV. The \text{ggF} and \text{VBF} samples are generated at next-to-leading-order (NLO) in QCD with \textsc{Powheg}~\cite{Alioli:2008tz,Nason:2009ai} with the parton showering modelled by \textsc{Pythia8}~\cite{Sjostrand:2007gs} for the 8 TeV samples and \textsc{Pythia}~\cite{pythia} for the 7~TeV samples.  The CT10~\cite{ct10} parton distribution functions (PDFs) are used in \textsc{Powheg}, with the ATLAS Underlying Event Tune, AU2~\cite{au2}. The \text{ggF} Higgs boson $\pt$ spectrum in \textsc{Powheg} is tuned to agree with the prediction from HRES~\cite{deFlorian:2012mx,Grazzini:2013mca}; 
this procedure shifts the Higgs boson $\pt$ spectrum to slightly smaller values.
The \text{VH} samples are generated with \textsc{Pythia8}, using AU2 and the CTEQ6~\cite{cteq6} PDFs. 

The predicted SM Higgs boson cross-sections and branching ratios are compiled in Refs.~\cite{LHCHiggsCrossSectionWorkingGroup:2011ti,LHCHiggsCrossSectionWorkingGroup:2012vm,Heinemeyer:2013tqa}. The cross section for the \text{ggF} process is calculated at next-to-next-to-leading-order (NNLO) ~\cite{Djouadi:1991tka,Dawson:1990zj,Spira:1995rr,Harlander:2002wh,Anastasiou:2002yz,Ravindran:2003um} in QCD and NLO electroweak (EW) corrections are applied~\cite{Aglietti2004,Actis2008}, assuming that the QCD and EW corrections factorise. The cross section for the \text{VBF} process is calculated with full NLO QCD and EW corrections ~\cite{Ciccolini:2007jr,Ciccolini:2007ec,Arnold:2008rz} and approximate NNLO QCD corrections~\cite{Bolzoni:2010xr}. The cross section for the \text{VH} process is calculated at NNLO~\cite{Han:1991ia,Brein:2003wg} in QCD, and NLO EW radiative corrections~\cite{Ciccolini:2003jy} are applied. The branching ratios for $\hmm$ decays as a function of $m_{H}$ are calculated using \textsc{Hdecay}~\cite{Djouadi:1997yw}.

The following MC event generators are used to simulate background processes: 
\textsc{Alpgen}~\cite{Mangano:2002ea}+\textsc{Herwig}~\cite{Corcella:2000bw} for $W$+jets, 
\textsc{MC@NLO}~\cite{mcatnlo}+\textsc{Herwig} for $\ttbar$, $tW$ and $tb$, 
\textsc{AcerMC}~\cite{Kersevan:2004yg}+\textsc{Pythia} for $tqb$, 
\textsc{Powheg}$+$\textsc{Pythia8} for $\zgam$ and $q\bar{q}\rightarrow WW$, 
gg2WW~\cite{gg2WW}+\textsc{Herwig} for $gg\rightarrow WW$, 
\textsc{Powheg}~\cite{Melia:2011tj}+\textsc{Pythia} for $WZ$ and $ZZ$ and 
\textsc{MadGraph}~\cite{Alwall:2007st,Alwall:2011uj}+\textsc{Pythia} for $W\gamma^{*}$. $WW$, $WZ$, $ZZ$ and $W\gamma^{*}$ production are referred to as diboson production later in the text.
In addition, contributions from $W$ boson production in association with one or more jets ($W$+jets), where one of the jets is misidentified as a muon, are estimated using a data control region containing same-sign dimuon events as described in Ref.~\cite{ATLAS:2013wqa}. 

The signal and background MC samples are processed through the ATLAS detector simulation~\cite{atlassim}  based on {\sc Geant4}~\cite{geant4}, 
followed by the same reconstruction algorithms that are used for collision data. The effects arising from multiple collisions in the same or neighbouring bunch crossings (pile-up) are included in the MC simulation, matching the pile-up conditions of the selected data sample.

\section{Event selection} \label{sec:event}
Events are required to contain at least one reconstructed $pp$ collision vertex candidate with at least three associated tracks each with $\pt>0.4$~GeV. The vertex with the largest sum of $p_{\rm T}^{\rm 2}$ of tracks is considered to be the primary vertex. Muon candidates are reconstructed by matching tracks in the inner detector to tracks reconstructed in the muon spectrometer. In addition to stringent track quality requirements imposed for muon identification, the muon tracks must be consistent 
with having originated from 
the primary vertex. All selected muon candidates are required to be within $|\eta| <$ 2.5. Muon candidates must pass track and calorimeter isolation requirements that scale with the $\pt$ of the muon track. The isolation is calculated as the scalar sum of the $\pt$ of additional tracks or the $E_{\rm T}$ of calorimeter energy deposits within cone of radius $\Delta R = \sqrt {(\Delta \phi)^2 + (\Delta \eta^2)} = 0.3$ around the muon track, normalised by $\pt^{\mu}$. 
For 8~TeV data, each muon with $\pt^{\mu}>20\ (15)$~GeV is required to have a normalised track isolation smaller than 0.12 (0.08) and a normalised calorimetric isolation less than 0.30 (0.18).
For 7~TeV data, equivalent values for $\pt^{\mu}>15$~GeV are 0.15 and 0.2 for track and
calorimetric isolation, respectively. Due to different pile-up
conditions, different isolation criteria are used for 7~TeV and 8~TeV data. 

Jets are reconstructed from clusters of calorimeter cells using the anti-$k_{t}$ algorithm~\cite{Cacciari:2008gp, Aad:2014bia} with a radius parameter of 0.4.
The selected jets must satisfy $E_{\rm T} > 25$ GeV for $|\eta| < 2.4$ and $E_{\rm T} > 30$ GeV for $2.4 \leq |\eta| < 4.5$. Muon candidates overlapping with the selected jets within a cone of radius $\Delta R = 0.4$ are removed from the analysis.
In the pseudorapidity range $|\eta| < 2.5$, jets originating from $b$-quarks 
are identified using a $b$-tagging algorithm~\cite{btag7tev, bperf1} 
with an efficiency of approximately 80\%, 
determined from $\ttbar$ MC events, and with a misidentification rate for selecting light-quark or gluon jets of less than 1\%.
The missing transverse momentum~\cite{Vichou:2012roa}, $E_{\rm T}^{\rm miss}$, is the magnitude of the vector sum of the $\pt$ of muons, electrons, photons, jets and clusters of calorimeter cells with $|\eta| < 4.9$ not associated with these objects.

Corrections are applied to simulated MC samples in order to account for differences between data and MC simulation for the trigger and identification efficiency and for the muon momentum scale and resolution.  The trigger and reconstruction efficiency corrections are measured using $Z\to\mumu$ events and are found to be within 2\% of unity. The muon momentum corrections are determined by comparing the reconstructed invariant mass distribution of $Z\to\mumu$ events in data with that from simulated events; these corrections are within 0.1\% of unity.  

\begin{figure}[h]
  \centering
  \includegraphics[width=0.4\textwidth]{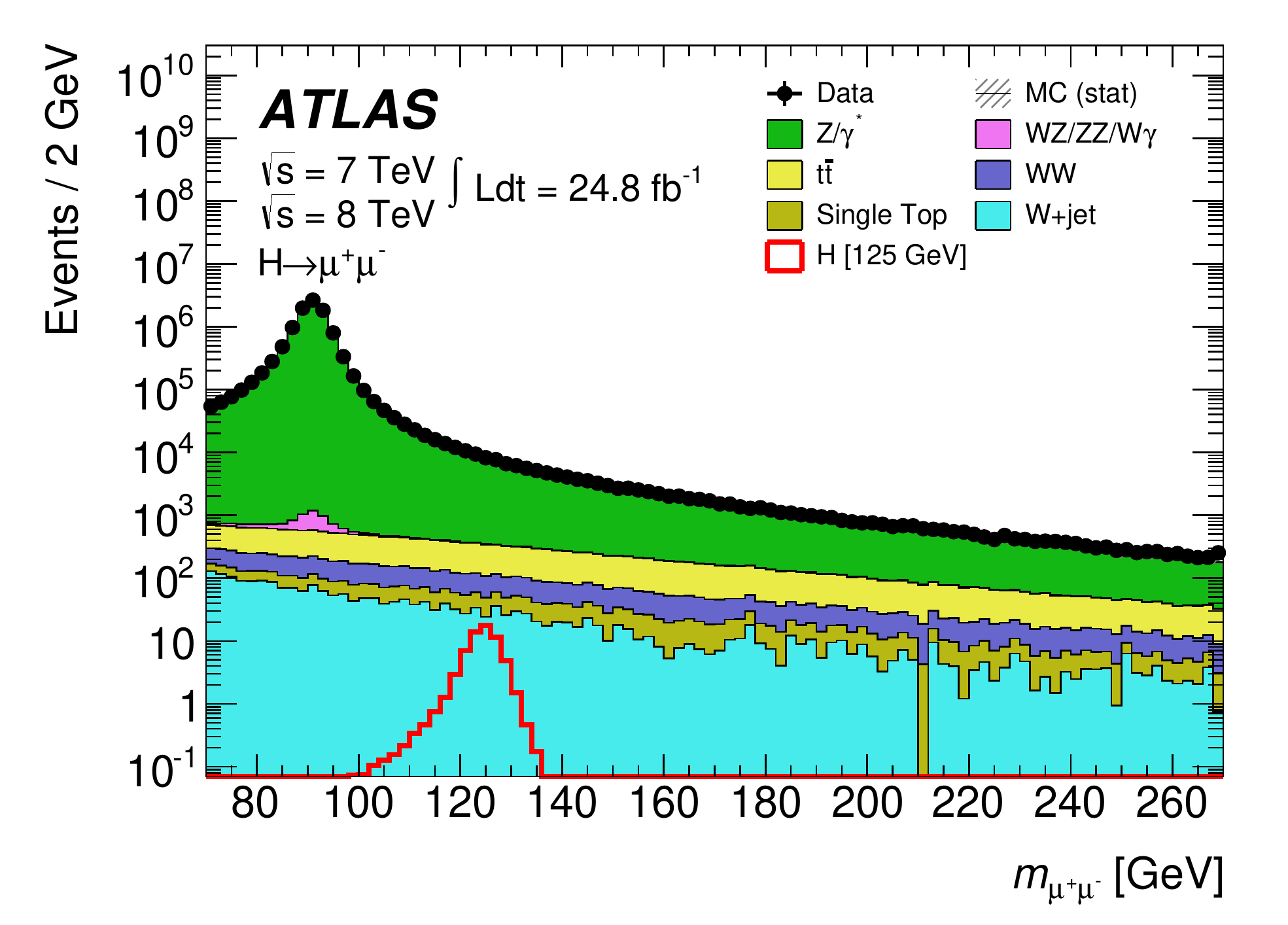}
  \includegraphics[width=0.4\textwidth]{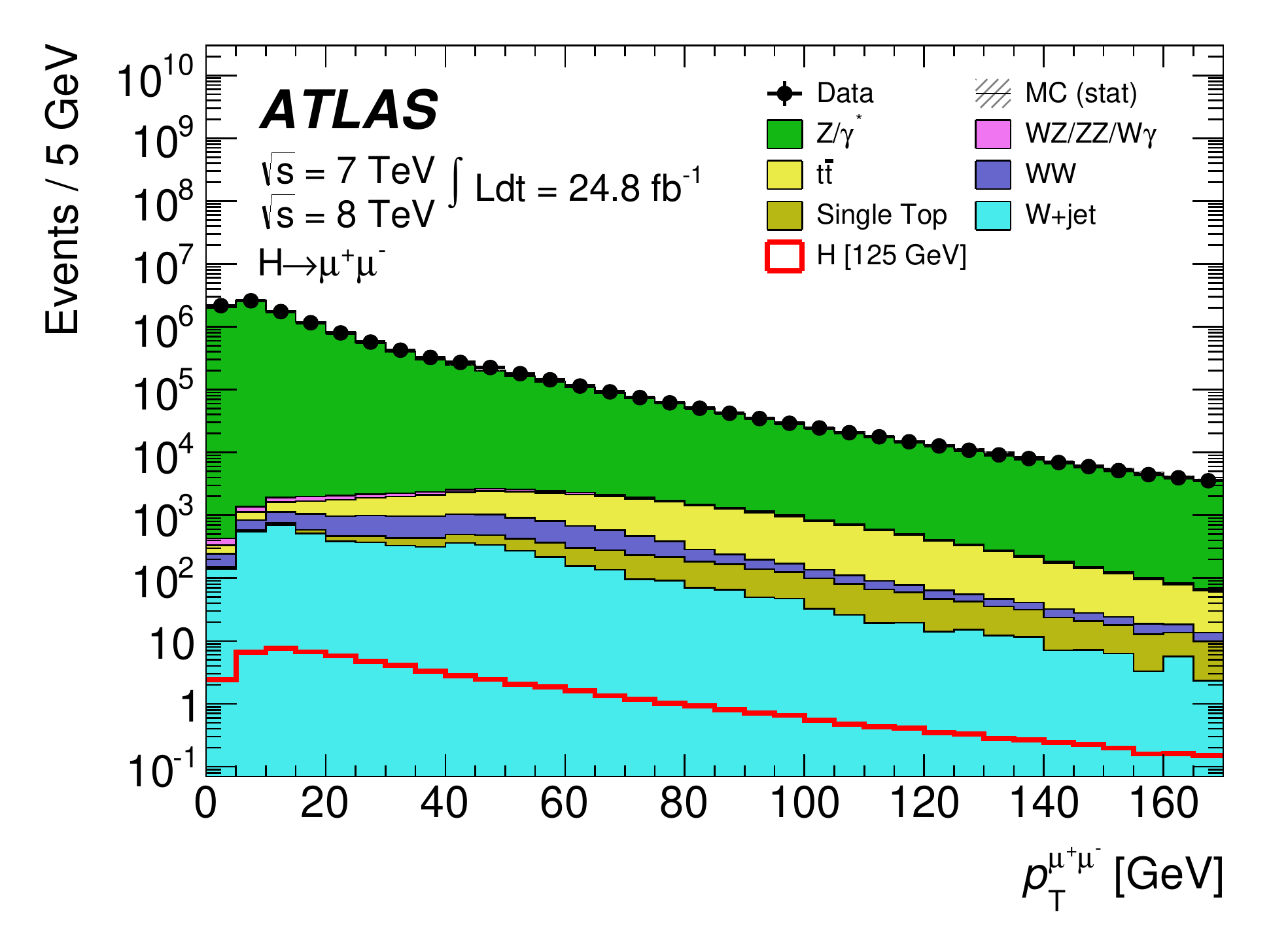}
  \caption{The distribution of 
    the dimuon invariant mass (top) and dimuon 
    momentum (bottom) for 7~TeV and 8~TeV data 
    with all the selection requirements described in Section~\ref{sec:event}. 
    The expected signal is shown for $m_{H}=125$~\GeV. 
    \label{fig:invmass-ptll}
  }
\end{figure}

$\hmm$ candidate events are selected by requiring exactly two oppositely charged muons with transverse momentum $\pt^{\mu_{1}}>25$~GeV and $\pt^{\mu_{2}}>15$~GeV for the leading and subleading muon, respectively. Selected events must contain at least one muon identified by the trigger system within a cone of radius $\Delta R = 0.15$ centred on the reconstructed muon candidate. The dominant background in this search is $\zgam \rightarrow \mu^+\mu^-$ production, followed by smaller backgrounds from single and pair production of top quarks and diboson processes. To suppress backgrounds from top quark pair production and diboson processes, events are required to have $E_{\rm T}^{\rm miss}<80$ GeV. The dimuon invariant mass distribution $m_{\mumu}$ and the dimuon transverse momentum \ptmm for data and MC events passing all the selection requirements so far are shown in Fig.~\ref{fig:invmass-ptll}. The number of expected signal events for $m_{H}=125$~GeV, the expected background contributions, and the number of observed data events in the $m_{\mumu}$ region from 122.5 to 127.5 GeV are shown in Table~\ref{tab:events}. The MC background yields are given to illustrate the expected background composition.
The selection efficiency times acceptance for signal events with 
$m_H = 125$~GeV after all selection criteria described thus far is approximately 55\%. 
%
%

The expected background processes produce smooth $m_{\mumu}$ distributions in the search window, allowing the total background normalisation and shape in each category to be derived from fitting the data as described in Section~\ref{sec:signalbg}. 
The $m_{\mumu}$ distribution is examined in the range \frange. This range is larger than the \hrange search window in order
 to account for signal resolution effects and to allow sufficient sidebands for background normalisation. 

\begin{table}[htbp]
\caption{\label{tab:events} Number of expected signal events for $m_{H}=125$~GeV, number of expected background events from MC simulation, and number of observed data events within a window of $|m_{H} - m_{\mumu}| \le 2.5$~GeV after all selection criteria are applied. Only statistical uncertainties are given. The theoretical systematic uncertainty on the dominant \zgam background is about 4\%. The MC background yields are given to illustrate the expected background composition.}
\centering
\vspace{0.2cm}
\scalebox{0.9}{
\begin{tabular}{lr@{$\,\pm \,$}l r@{$\,\pm \,$}l }
\hline\hline
& \multicolumn{2}{c}{7~TeV} & \multicolumn{2}{c}{8~TeV}  \\
\hline
Signal (125 GeV)   & 5.6  & 0.1  &  32.7 & 0.2\\
\hline                                       
\zgam              & 3110 & 40   & 16660 & 270\\
$WZ/ZZ/W\gamma$    & 2.2  & 0.2  & 12.3  & 0.7\\
$t\bar{t}$         & 75.6 & 1.8  & 509.2 & 2.7\\
$WW$               & 23.2 & 0.5  & 123.3 & 1.6\\
Single top         & 7.2  & 0.9  & 54.5  & 0.6\\
$W$+jets           & 3.2  & 1.5  & 38  & 4  \\
\hline                                      
Total Bkg.         & 3220 & 40   & 17390 & 270\\
\hline 
Observed          & \multicolumn{2}{c}{3344} & \multicolumn{2}{c}{17745} \\
\hline\hline
\end{tabular}
}
\end{table}

\section{Event categorisation}\label{sec:categorisation}

To increase sensitivity to the Higgs boson signal, the selected events are separated into seven mutually exclusive categories with different signal-to-background ratios based on their muon pseudorapidity ($\eta_{\mu}$), \ptmm , and \text{VBF} dijet signature. Events produced in the \text{VBF} process are characterised by two forward jets with little hadronic activity between them. The \text{VBF} category is thus defined by requiring the events to have at least two jets with an invariant mass greater than 500 GeV, $|\eta_{\text{jet}_{1}} - \eta_{\text{jet}_{2}}| > 3$ and $\eta_{\text{jet}_{1}}\times\eta_{\text{jet}_{2}}<0$. In events with more than two jets, those with the highest $\pt$ are used in the selection. Events with at least one jet identified as originating from a $b$-quark are excluded from the VBF category.

The events that are not selected for the \text{VBF} category are classified using \ptmm. Signal events have on average larger values of \ptmm than the $\zgam$ background events. Therefore, the remaining events are separated into three \ptmm categories: low ($<15$~GeV), medium (15--50~GeV) and high ($>50$~GeV). To further improve the search sensitivity, each of these three categories is also subdivided into a central category with $|\eta_{\mu_{1}}|<1$ and $|\eta_{\mu_{2}}|<1$ and a non-central category containing all remaining events. This value for the $\eta_{\mu}$ boundary has been chosen by scanning a range of $\eta_{\mu}$ values and selecting a value with the highest signal sensitivity. The muon momentum measurement for the central muons is more precise, producing a narrower $m_{\mumu}$ distribution for signal events in the central category and thus resulting in a higher overall signal sensitivity. 
Table~\ref{tab:categories} shows the signal event yields, $N_{\rm S}/\sqrt{N_{\rm B}}$ ratios, approximate signal width 
and results of the fits to the data, described in Section~\ref{sec:signalbg}, for all analysis categories.
\begin{table}[h]
  \begin{small}
    \setlength{\tabcolsep}{4pt}
    \caption{Expected signal yields ($N_{\rm S}$) for $m_H=125$ GeV and the ratio 
      $N_{\rm S}/\sqrt{N_{\rm B}}$ using the simulated MC background yields 
      ($N_{\rm B}$) within a window of $|m_{H} - m_{\mumu}| \le 2.5$~GeV 
      for each of the event categories under study.
      In addition, the full width at half maximum (FWHM) of the signal 
      $m_{\mumu}$ distribution, modelled as
      described in Section~\ref{sec:signalbg}, is given. 
      Also shown are $\chi^2/\text{ndof}$ of the fits to 
      the $m_{\mumu}$ distribution described in Section~\ref{sec:signalbg}.
    }
    \label{tab:categories}
    \begin{center}
      \scalebox{0.88}{
      \begin{tabular}{llcrrr}
        \hline\hline\\
        $\sqrt{s}$ & Category & $N_{\rm S}$ & $\frac{N_{\rm S}}{\sqrt{N_{\rm B}}}$ & FWHM & $\chi^2/\text{ndof}$ \\
        $[$TeV$]$  &          &            &                                  & [GeV] & \\
        \hline
        8 & non-cen. low \ptmm                & 6.1  & 0.07 & 6.6 & 49.8/48  \\
        8 & cen. low \ptmm                    & 2.6  & 0.06 & 5.5 & 52.8/48  \\
        8 & non-cen. medium \ptmm             & 10.4 & 0.15 & 6.6 & 45.1/48 \\
        8 & cen. medium \ptmm                 & 4.7  & 0.13 & 5.6 & 36.7/48 \\
        8 & non-cen. high \ptmm               & 5.5  & 0.13 & 7.2 & 26.7/48 \\
        8 & cen. high \ptmm                   & 2.6  & 0.10 & 6.0 & 32.3/48 \\
        8 & VBF                               & 0.8  & 0.09 & 7.0 & 18.6/19 \\
        \hline
        7 & non-cen. low \ptmm                & 1.0 & 0.03 & 6.8 & 42.0/48 \\
        7 & cen. low \ptmm                    & 0.5 & 0.03 & 5.3 & 43.5/48 \\
        7 & non-cen. medium \ptmm             & 1.8 & 0.06 & 6.9 & 41.2/48 \\
        7 & cen. medium \ptmm                 & 0.8 & 0.05 & 5.5 & 34.4/48 \\
        7 & non-cen. high \ptmm               & 0.9 & 0.05 & 7.5 & 60.0/48 \\
        7 & cen. high \ptmm                   & 0.5 & 0.05 & 5.9 & 56.2/48 \\
        7 & VBF                               & 0.1 & 0.05 & 6.9 & 6.2/19 \\
        \hline\hline
      \end{tabular}
    }
    \end{center}
  \end{small}
\end{table}

\section{Signal and background models}\label{sec:signalbg}
Analytical models are used to describe the $m_{\mumu}$ distributions for signal and background processes. 
The simulated samples detailed in Section~\ref{sec:data} are used to develop background models
which are designed to describe essential features of the background $m_{\mumu}$ distributions, dominated 
by \zgam, while having sufficient flexibility to describe different categories and to absorb potential differences between data and MC simulation.

\begin{figure}[h]
  \centering
  \includegraphics[width=0.38\textwidth]{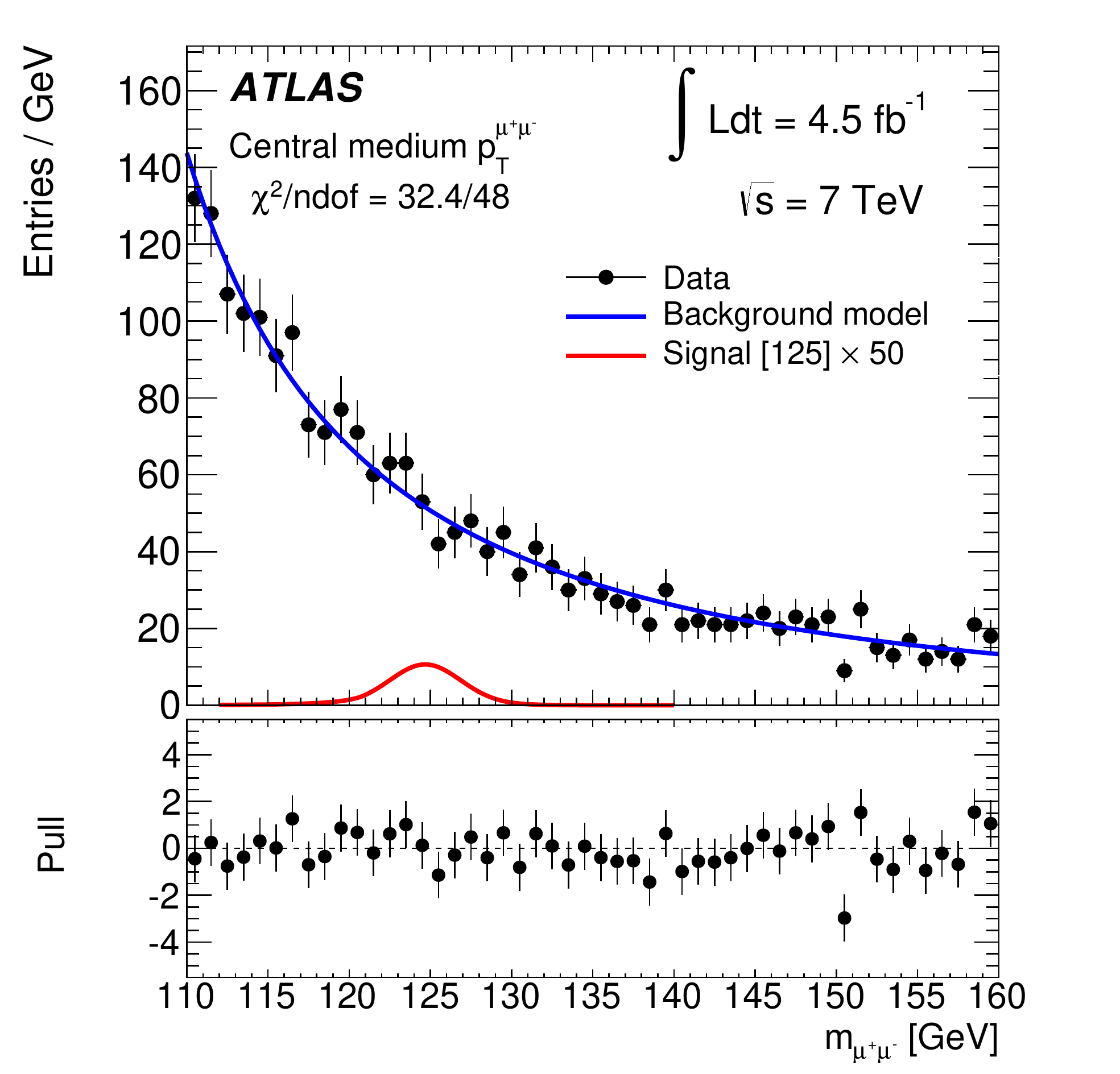}
  \includegraphics[width=0.38\textwidth]{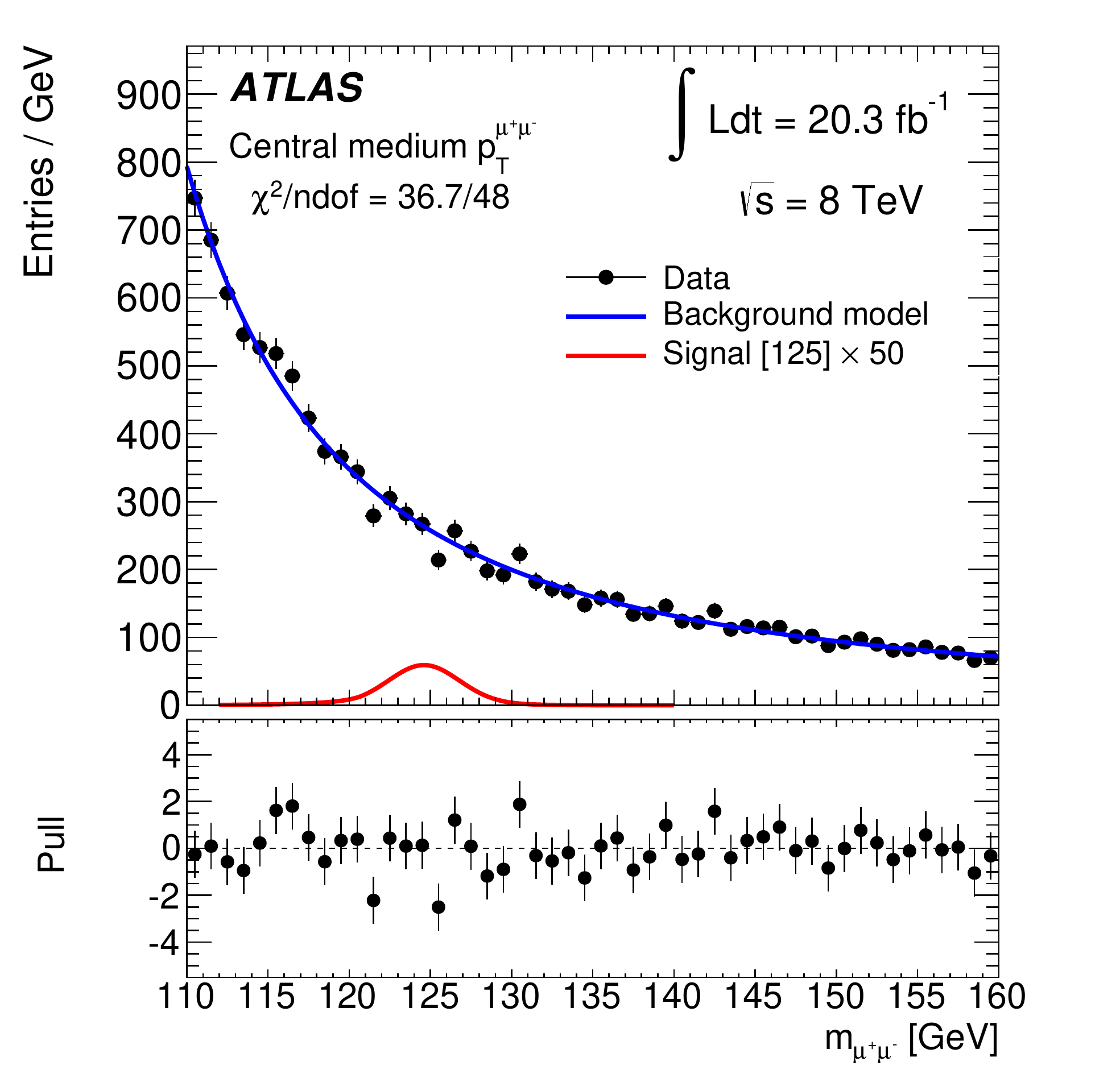}
  \caption{The background model fits to the $m_{\mumu}$ distribution 
    for the central medium \ptmm category for 7~TeV (top) and 8~TeV (bottom) data. 
    The statistical uncertainties are given for the data points.
    The expected signal is shown for $m_{H}=125$~\GeV\ and is scaled by a factor of 50.
    \label{fig:fit}
  }
\end{figure}

The background model selected for the $\ptmm$ categories is 
the sum of a Breit--Wigner (BW) function convolved with a Gaussian 
function (GS), and an exponential function divided by $x^{3}$:
\begin{equation}
\begin{split}
  \label{equ:bg}
P_{\rm B}(x) = & f \cdot [{\rm BW}(M_{\rm BW}, \Gamma_{\rm BW}) \ast {\rm GS}(\sigma_{\rm GS}^{\rm B})](x) \\
&+ (1-f) \cdot C \cdot {\rm e}^{A\cdot x}/x^{3},
\end{split}
\end{equation} 
where $x$ represents $m_{\mumu}$ and $f$ represents the fraction of the BW component 
when each individual component is normalised to unity. $C$ is an overall normalisation coefficient. The $\sigma_{\rm GS}^{\rm B}$ parameters in each category 
are fixed to the average $m_{\mumu}$ resolution in that category as determined from the MC simulation of \zgam. 
The background model for the \text{VBF} category is the product of a Breit--Wigner and an exponential function:
\begin{equation}
  \label{equ:vbf}
P_{\rm B}(x) = {\rm BW}(M_{\rm BW}, \Gamma_{\rm BW}, x) \cdot {\rm e}^{A\cdot x}.
\end{equation}
For all categories, the BW parameters are fixed to $M_{\rm BW}=91.2$~GeV and $\Gamma_{\rm BW} = 2.49$~GeV. The parameters $f$, $A$ and the overall background normalisation are determined from fits to the data, as shown in Fig.~\ref{fig:fit} for the central medium \ptmm category. Similar fit quality is observed for all other categories.
%
%

The signal model is obtained from simulated Higgs boson signal samples, where contributions from the \text{ggF}, \text{VBF} and \text{VH} Higgs boson production processes are added together. This model is the sum of a Crystal Ball (CB)\footnote{A Gaussian function with a power-law tail.} and a Gaussian function:
\begin{equation}
  \label{eq:signal}
\begin{split}
  P_{\rm S}(x) = & f_{\rm CB} \cdot {\rm CB}(x, m, \sigma_{\rm CB}, \alpha, n) \\
  & + (1-f_{\rm CB})\cdot {\rm GS} \left(x, m, \sigma_{\rm GS}^{\rm S}\right),
\end{split}
\end{equation} 
where $x$ represents $m_{\mumu}$ and $f_{\rm CB}$ represents the fraction of the CB contribution when each individual component is normalised to unity. 
The parameters $\alpha$ and $n$ define the power-law tail
of the CB distribution.
The parameters $\sigma_{\rm CB}$ and $\sigma_{\rm GS}^{\rm S}$ denote the widths of the CB and GS distributions, respectively. The parameters $m$, $\sigma_{\rm CB}$ and $\sigma_{\rm GS}^{\rm S}$ 
are determined from the fits to the simulated Higgs boson samples.
In order to improve stability of the fits, the remaining parameters $f_{\rm CB}$, $\alpha$ and $n$
are fixed to values determined from empirical tests where a range of possible values have been tested.

\begin{figure}[htbp]
   \centering
   \includegraphics[width=0.4\textwidth]{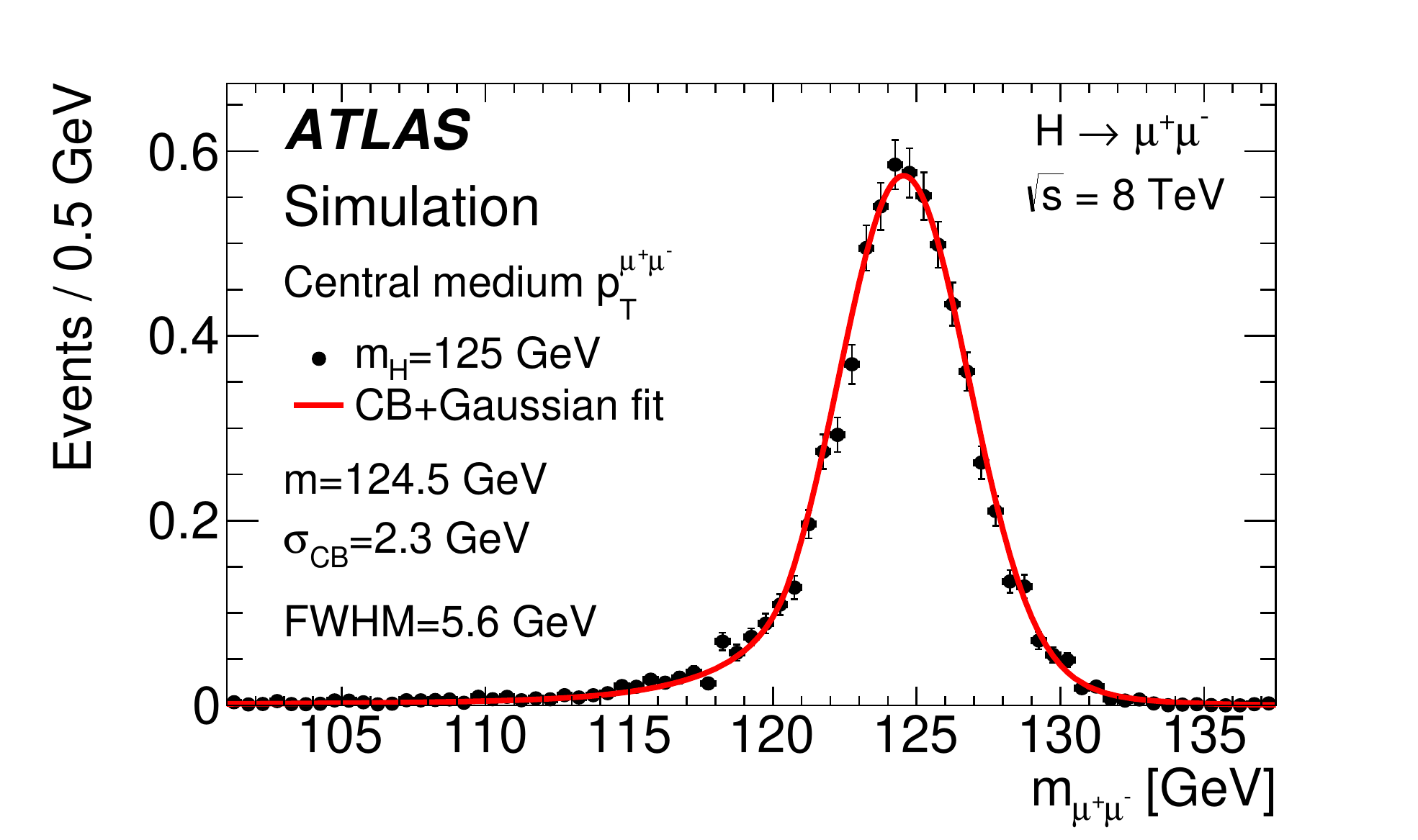}\\
   \includegraphics[width=0.4\textwidth]{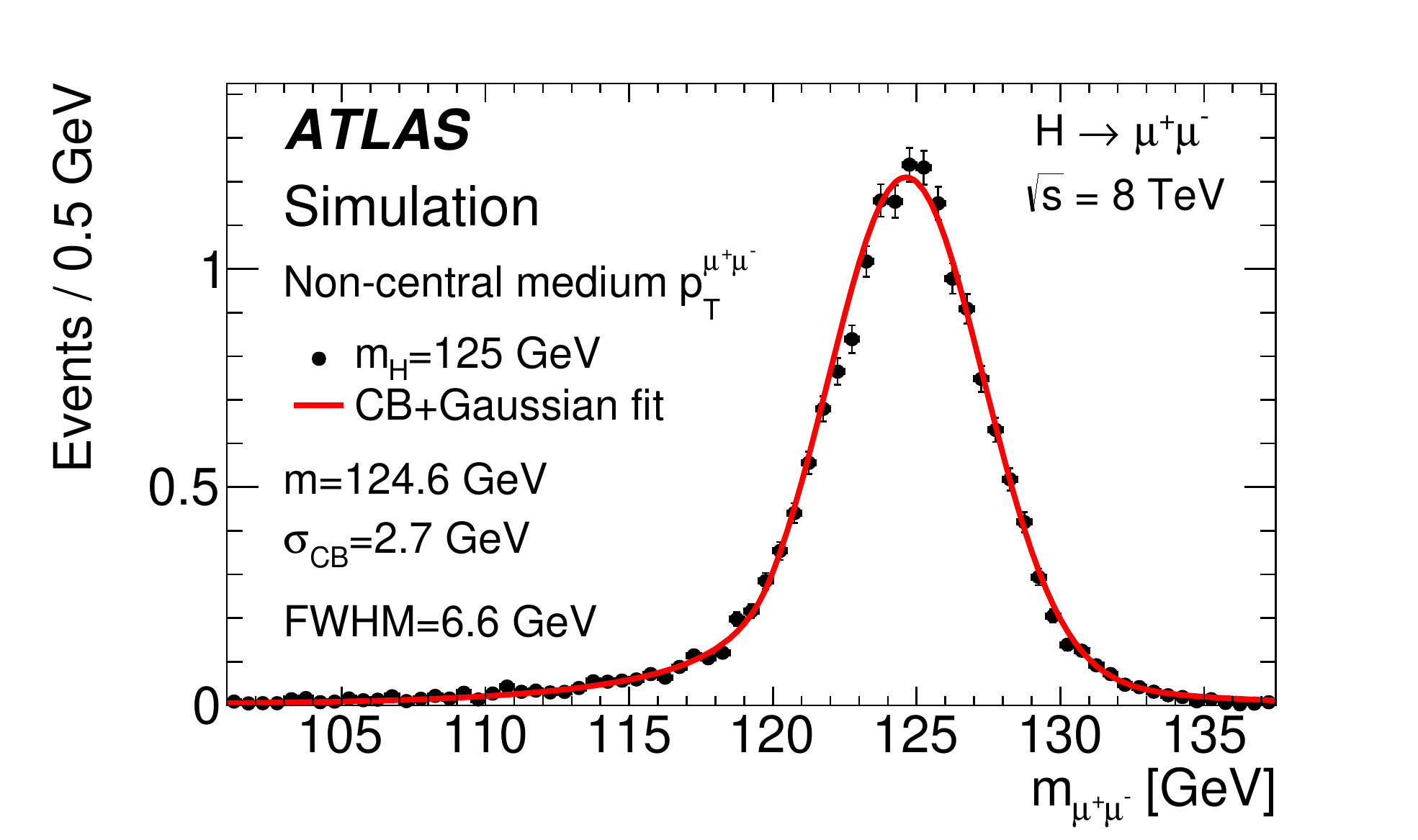}
   \caption{The signal model fit to the $m_{\mumu}$ distribution for 
     the central (top) and non-central (bottom) simulated Higgs boson events for $m_{H}=125$~GeV 
     in the medium \ptmm category for $\sqrt{s}=8$~TeV.
     \label{fig:fitsig}
   }
\end{figure}

Fig.~\ref{fig:fitsig} shows how the signal model reproduces the simulation 
for the medium \ptmm category
for the expected signal dimuon mass distributions. Similar fit quality is obtained for all other categories.
The signal model parameters are linearly interpolated in steps of 1~GeV between the generated signal samples. These signal model parameters are then fixed in the fit to data.

%

To derive the results presented in Section~\ref{sec:results}, a binned maximum likelihood fit to the observed $m_{\mumu}$ distributions 
in the range \frange is performed using the sum of the signal and background model. The fit is done simultaneously in all seven categories 
with separate distributions for 7~TeV and 8~TeV data. 
In addition to the background model parameters described earlier, 
a background normalisation in each category and the \hmm signal strength $\mu_{\rm S}$ 
are free parameters of the fit, where $\mu_{\rm S}$ is defined such that $\mu_{\rm S}=0$ 
corresponds to the background-only hypothesis 
and $\mu_{\rm S}=1$ corresponds to the SM $\hmm$ signal.

\section{Systematic uncertainties}\label{sec:systematics}
The main theoretical and experimental sources of uncertainty on the number of 
expected signal events are shown in Table~\ref{tab:systexp}. 
The uncertainty on the integrated luminosity is $\pm1.8\%$ for 7~TeV data~\cite{Aad:2013ucp} and 
$\pm2.8\%$ for 8~TeV data; 
it is obtained following the same methodology 
as that detailed in Ref.~\cite{Aad:2013ucp}, from a preliminary calibration 
of the luminosity scale derived from beam-separation scans performed in November 2012. 

Sources of experimental uncertainty include the efficiency of the muon trigger, reconstruction, identification, and isolation requirements, as well as the muon momentum scale and resolution.
Uncertainties on the jet energy scale and resolution affect the 
selection of jets used in the \text{VBF} category definitions. Smaller uncertainties
arise from pile-up and the primary vertex selection.

The theoretical uncertainties on the production and $\hmm$ decay of a SM Higgs boson of mass
 $m_H=125$~GeV are taken from Refs.~\cite{LHCHiggsCrossSectionWorkingGroup:2011ti,LHCHiggsCrossSectionWorkingGroup:2012vm}. 
The uncertainty on the relative populations of the \ptmm\ categories, due
to the uncertainty on the description of the Higgs boson \pt\ spectrum 
arising from missing higher-order corrections, is determined by varying 
the QCD renormalisation, factorisation and resummation scales used in the HRES program.
To evaluate these uncertainties, the scales are independently varied 
up and down by a factor of two while keeping their ratio between 0.5 and 2.0.
The \text{ggF} contribution to the \text{VBF} category has large 
uncertainties due to missing higher-order corrections; they are estimated 
using the method described in Ref.~\cite{Heinemeyer:2013tqa}. 
The uncertainties associated with the modelling of multi-parton interactions (MPI) 
 are estimated by turning off the MPI modelling in the event generation,
according to the recommendations in Ref.~\cite{Heinemeyer:2013tqa}.

In addition to the samples described in Section~\ref{sec:data}, samples of the dominant $\zgam$ background are generated with \textsc{Powheg}$+$\textsc{Pythia8} and parameterised with a detector response measured using simulated MC events. These samples contain approximately 170 times more events than expected in the data and are used to validate the background models and to derive systematic uncertainties due to potential mismodelling bias. This bias is estimated by fitting  the parametrised signal plus background model to the simulated $m_{\mu^+\mu^-}$ background distribution in the mass range \frange where the signal strength $\mu_{\rm S}$ is a free parameter.
The bias is then defined as the root mean square of the signal yield obtained
from the fit for Higgs boson masses in the range
\hrange. This uncertainty varies from 3\% to 20\% of the statistical uncertainty 
on the signal strength $\mu_{\rm S}$,
depending on the selection category and data-taking period.

\begin{table}[h]
  \centering
  \caption{\label{tab:systexp} Main sources of experimental 
    and theoretical uncertainty on the signal yield, excepting the error 
    from mismodelling bias.
    ``QCD scale'' indicates the theoretical uncertainty on the Higgs 
    boson production due to missing higher-order corrections estimated 
     by varying the QCD renormalisation  
     and factorisation scales, while ``PDFs + $\alpha_{s}$'' indicates 
     uncertainty due to parton distribution functions, as described in Ref.~\cite{LHCHiggsCrossSectionWorkingGroup:2011ti,LHCHiggsCrossSectionWorkingGroup:2012vm}. 
     The ranges for the uncertainties cover the variations 
     among different categories and data-taking periods.
   }
   \vspace{0.2cm}
   \scalebox{0.85}{
  \begin{tabular}{ll}
    \hline\hline
    Source (experimental)       & Uncertainty (\%) \\
    \quad  Luminosity           & $\pm$1.8 (7~TeV), $\pm$2.8 (8~TeV) \\ 
    \quad Muon efficiency       & $\pm$1 \\
    \quad Muon momentum res.        & $\pm$1  \\ 
    \quad Muon trigger          & $\pm$1.5   \\
    \quad Muon isolation        & $\pm$1.1   \\
    \quad Pile-up reweighting   & $\pm$1   \\
    \quad Jet energy scale      & $^{+3.4}_{-4.5}$ (\text{VBF})\\[0.1cm]
    \hline
    Source (theory) & Uncertainty (\%) \\
    \quad Higgs boson branching ratio       & $\pm$7 \\
    \quad QCD scale  &  $\pm8$ (\text{ggF}), $\pm1$ (\text{VBF}, \text{VH})\\
    \quad PDFs + $\alpha_{\rm s}$    &  $\pm 8$  (\text{ggF}), $\pm 4$ (\text{VBF}, \text{VH})\\ 
    \quad \text{ggF} uncert. in \text{VBF}    & $\pm$22 \\
    \quad Multi-parton inter. in \text{VBF} & $\pm 9$ (\text{ggF}), $\pm 4$ (\text{VBF}) \\
    \hline\hline
  \end{tabular}
}
\end{table}

\section{Results and conclusions} \label{sec:results}
The statistical procedure used to interpret the data is summarised in Ref.~\cite{ATLAS:2012gk}. The observed data is consistent with the expected backgrounds
and no evidence for a signal is found. Upper limits are computed on the signal strength $\mu_{\rm S}$ using a modified frequentist CL$_\text{s}$ method~\cite{Cowan:2010js,Read:2002hq} based on a Poisson log-likelihood ratio statistical test.

\begin{figure}[htbp]
\centering
  \includegraphics[width=0.4\textwidth]{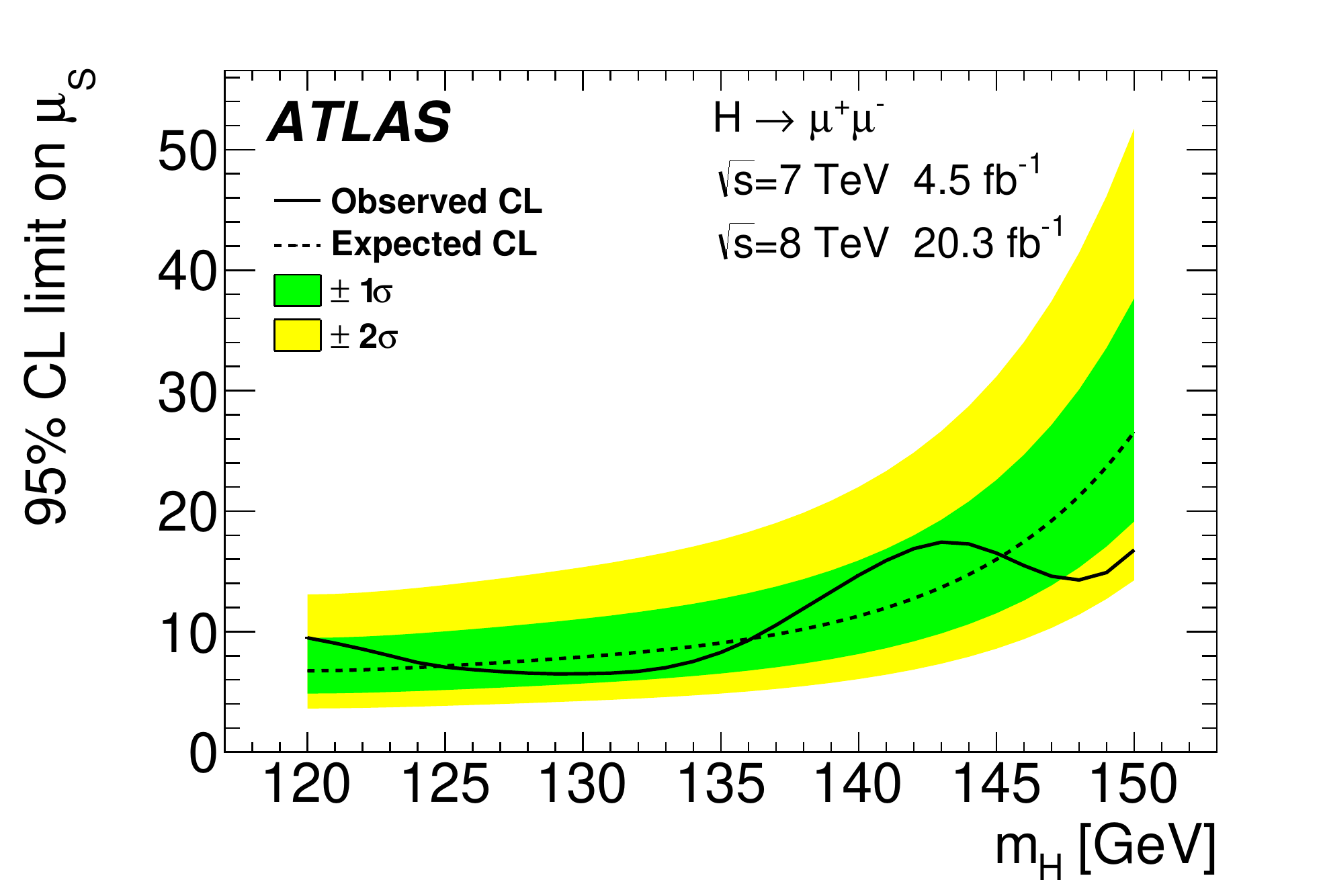}
  \caption{Observed (solid) and expected (dashed) 95\% CL upper limits on the \hmm signal strength as a function of $m_H$ over the mass range \hrange. The dark- and light-shaded regions indicate the $\pm 1\sigma$ and $\pm2\sigma$ uncertainty bands on the expected limit, respectively. \label{fig:f-limits}}
\end{figure}

The observed and expected 95\% confidence level (CL) limits on the \hmm signal strength are shown in Fig.~\ref{fig:f-limits}. 
Table~\ref{tab:limits} summarises the observed and expected limits for different values of $m_H$. 
Including the systematic uncertainties described in Section~\ref{sec:systematics} changes
the expected limit by approximately 2\% for $m_{H}=125$~GeV.
%

%
\begin{table}[htbp]
 \begin{center}
   \caption{Observed and expected 95\% CL upper limits on the \hmm signal strength $\mu_{\rm S}$ for different values of $m_H$.
   \label{tab:limits}
  }
  \vspace{0.2cm}
 \scalebox{0.85}{
 \begin{tabular}{l|rrrrrrr}
\hline
\hline
$m_{H}$ [GeV] & 120 & 125 & 130 & 135 & 140 & 145 & 150 \\
\hline
Obs.         & 9.5 & 7.1 & 6.5 & 8.3 & 14.7 & 16.5 & 16.8\\
Exp.         & 6.7 & 7.2 & 7.9 & 9.1 & 11.3 & 16.0 & 26.6\\
\hline           
\hline           
\end{tabular}
}
 \end{center}
\end{table}

To conclude, a search for Higgs boson decay to $\mu^{+}\mu^{-}$ in 
\lumi of $pp$ collisions at $\sqrt{s}=$7 and 8~TeV at the LHC has been 
performed with the ATLAS experiment. The observed data is consistent with the expected 
backgrounds. 
No evidence for a signal is observed and upper limits are set on the signal strength as a function of the Higgs boson mass. 
For a SM Higgs boson with a mass of 125.5~GeV, the observed (expected) limit on the signal strength $\mu_{\rm S}$ at the 95\% CL is 7.0 (7.2) times the SM prediction. 
Assuming a Higgs boson mass of $125.5$~GeV and the SM production cross section, which is allowed to vary within its uncertainty, 
the 95\% CL upper limit on the \hmm branching ratio is $1.5\times10^{-3}$.


\section*{Acknowledgements}

We thank CERN for the very successful operation of the LHC, as well as the
support staff from our institutions without whom ATLAS could not be
operated efficiently.

We acknowledge the support of ANPCyT, Argentina; YerPhI, Armenia; ARC,
Australia; BMWF and FWF, Austria; ANAS, Azerbaijan; SSTC, Belarus; CNPq and FAPESP,
Brazil; NSERC, NRC and CFI, Canada; CERN; CONICYT, Chile; CAS, MOST and NSFC,
China; COLCIENCIAS, Colombia; MSMT CR, MPO CR and VSC CR, Czech Republic;
DNRF, DNSRC and Lundbeck Foundation, Denmark; EPLANET, ERC and NSRF, European Union;
IN2P3-CNRS, CEA-DSM/IRFU, France; GNSF, Georgia; BMBF, DFG, HGF, MPG and AvH
Foundation, Germany; GSRT and NSRF, Greece; ISF, MINERVA, GIF, I-CORE and Benoziyo Center,
Israel; INFN, Italy; MEXT and JSPS, Japan; CNRST, Morocco; FOM and NWO,
Netherlands; BRF and RCN, Norway; MNiSW and NCN, Poland; GRICES and FCT, Portugal; MNE/IFA, Romania; MES of Russia and ROSATOM, Russian Federation; JINR; MSTD,
Serbia; MSSR, Slovakia; ARRS and MIZ\v{S}, Slovenia; DST/NRF, South Africa;
MINECO, Spain; SRC and Wallenberg Foundation, Sweden; SER, SNSF and Cantons of
Bern and Geneva, Switzerland; NSC, Taiwan; TAEK, Turkey; STFC, the Royal
Society and Leverhulme Trust, United Kingdom; DOE and NSF, United States of
America.

The crucial computing support from all WLCG partners is acknowledged
gratefully, in particular from CERN and the ATLAS Tier-1 facilities at
TRIUMF (Canada), NDGF (Denmark, Norway, Sweden), CC-IN2P3 (France),
KIT/GridKA (Germany), INFN-CNAF (Italy), NL-T1 (Netherlands), PIC (Spain),
ASGC (Taiwan), RAL (UK) and BNL (USA) and in the Tier-2 facilities
worldwide.

\section*{References}
\bibliographystyle{atlasBibStyleWithTitle}
\bibliography{main}

\providecommand{\href}[2]{#2}\begingroup\raggedright\begin{thebibliography}{10}

\bibitem{Englert:1964et}
F.~Englert and R.~Brout, {\em {Broken symmetry and the mass of gauge vector
  mesons}},
\href{http://dx.doi.org/10.1103/PhysRevLett.13.321}{Phys. Rev. Lett. {\bfseries
  13} (1964) 321--323}.

\bibitem{Higgs:1964pj}
P.~W. Higgs, {\em {Broken symmetries and the masses of gauge bosons}},
\href{http://dx.doi.org/10.1103/PhysRevLett.13.508}{Phys. Rev. Lett. {\bfseries
  13} (1964) 508--509}.

\bibitem{Guralnik:1964eu}
G.~Guralnik, C.~Hagen, and T.~Kibble, {\em {Global conservation laws and
  massless particles}},
\href{http://dx.doi.org/10.1103/PhysRevLett.13.585}{Phys. Rev. Lett. {\bfseries
  13} (1964) 585--587}.

\bibitem{Weinberg:1967tq}
S.~Weinberg, {\em {A Model of Leptons}},
\href{http://dx.doi.org/10.1103/PhysRevLett.19.1264}{Phys. Rev. Lett.
  {\bfseries 19} (1967) 1264--1266}.

\bibitem{Glashow:1961tr}
S.~L. Glashow, {\em {Partial Symmetries of Weak Interactions}},
\href{http://dx.doi.org/10.1016/0029-5582(61)90469-2}{Nucl. Phys. {\bfseries
  22} (1961) 579--588}.

\bibitem{Salam:1968rm}
A.~Salam, {\em {Weak and Electromagnetic Interactions}}, Conf.Proc. {\bfseries
  C 680519} (1968) 367--377.

\bibitem{ATLAS:2012gk}
{ATLAS} Collaboration, {\em {Observation of a new particle in the search for
  the Standard Model Higgs boson with the ATLAS detector at the LHC}},
  \href{http://dx.doi.org/10.1016/j.physletb.2012.08.020}{Phys. Lett.
  {\bfseries B 716} (2012) 1--29},
\href{http://arxiv.org/abs/1207.7214}{{\ttfamily arXiv:1207.7214}}.

\bibitem{CMS:2012gu}
{CMS} Collaboration, {\em {Observation of a new boson at a mass of 125 GeV with
  the CMS experiment at the LHC}},
  \href{http://dx.doi.org/10.1016/j.physletb.2012.08.021}{Phys. Lett.
  {\bfseries B 716} (2012) 30--61},
\href{http://arxiv.org/abs/1207.7235}{{\ttfamily arXiv:1207.7235}}.

\bibitem{ATLAS:2013wqa}
{ATLAS} Collaboration, {\em {Measurements of Higgs boson production and
  couplings in diboson final states with the ATLAS detector at the LHC}},
  \href{http://dx.doi.org/10.1016/j.physletb.2013.08.010}{Phys. Lett.
  {\bfseries B 726} (2013) 88--119},
\href{http://arxiv.org/abs/1307.1427}{{\ttfamily arXiv:1307.1427}}.

\bibitem{ATLAS:2013xqa}
{ATLAS} Collaboration, {\em {Evidence for the spin-0 nature of the Higgs boson
  using ATLAS data}},
  \href{http://dx.doi.org/10.1016/j.physletb.2013.08.026}{Phys. Lett.
  {\bfseries B 726} (2013) 120--144},
\href{http://arxiv.org/abs/1307.1432}{{\ttfamily arXiv:1307.1432}}.

\bibitem{CMS:2014vua}
{CMS} Collaboration, {\em {Evidence for the direct decay of the 125 GeV Higgs
  boson to fermions}},
\href{http://arxiv.org/abs/1401.6527}{{\ttfamily arXiv:1401.6527}}.

\bibitem{CMS:2013lba}
{CMS} Collaboration, {\em {Observation of a new boson with mass near 125 GeV in
  pp collisions at $\sqrt{s}$ = 7 and 8 TeV}},
  \href{http://dx.doi.org/10.1007/JHEP06(2013)081}{JHEP {\bfseries 1306} (2013)
  081},
\href{http://arxiv.org/abs/1303.4571}{{\ttfamily arXiv:1303.4571}}.

\bibitem{LHCHiggsCrossSectionWorkingGroup:2011ti}
{LHC Higgs Cross Section Working Group}, S.~Dittmaier, C.~Mariotti,
  G.~Passarino, and R.~Tanaka~(Eds.), {\em {Handbook of LHC Higgs Cross
  Sections: 1. Inclusive Observables}}, CERN-2011-002 (CERN, Geneva, 2011),
\href{http://arxiv.org/abs/1101.0593}{{\ttfamily arXiv:1101.0593}}.

\bibitem{LHCHiggsCrossSectionWorkingGroup:2012vm}
{LHC Higgs Cross Section Working Group}, S.~Dittmaier, C.~Mariotti,
  G.~Passarino, and R.~Tanaka~(Eds.), {\em {Handbook of LHC Higgs Cross
  Sections: 2. Differential Distributions}}, CERN-2012-002 (CERN, Geneva,
  2012),
\href{http://arxiv.org/abs/1201.3084}{{\ttfamily arXiv:1201.3084}}.

\bibitem{atlas-det}
{ATLAS} Collaboration, {\em {The ATLAS experiment at the CERN Large Hadron
  Collider}},
\href{http://dx.doi.org/10.1088/1748-0221/3/08/S08003}{JINST {\bfseries 3}
  (2008) S08003}.

\bibitem{Aad:2012xs}
{ATLAS} Collaboration, {\em {Performance of the ATLAS trigger system in 2010}},
  \href{http://dx.doi.org/10.1140/epjc/s10052-011-1849-1}{Eur. Phys. J.
  {\bfseries C 72} (2012) 1849},
\href{http://arxiv.org/abs/1110.1530}{{\ttfamily arXiv:1110.1530}}.

\bibitem{Alioli:2008tz}
S.~Alioli, P.~Nason, C.~Oleari, and E.~Re, {\em {NLO Higgs boson production via
  gluon fusion matched with shower in POWHEG}},
  \href{http://dx.doi.org/10.1088/1126-6708/2009/04/002}{JHEP {\bfseries 0904}
  (2009) 002}, \href{http://arxiv.org/abs/0812.0578}{{\ttfamily
  arXiv:0812.0578}}.

\bibitem{Nason:2009ai}
P.~Nason and C.~Oleari, {\em {NLO Higgs boson production via vector-boson
  fusion matched with shower in POWHEG}},
  \href{http://dx.doi.org/10.1007/JHEP02(2010)037}{JHEP {\bfseries 1002} (2010)
  037},
\href{http://arxiv.org/abs/0911.5299}{{\ttfamily arXiv:0911.5299}}.

\bibitem{Sjostrand:2007gs}
T.~Sjostrand, S.~Mrenna, and P.~Z. Skands, {\em {A brief introduction to PYTHIA
  8.1}}, \href{http://dx.doi.org/10.1016/j.cpc.2008.01.036}{Comput. Phys.
  Commun. {\bfseries 178} (2008) 852--867},
\href{http://arxiv.org/abs/0710.3820}{{\ttfamily arXiv:0710.3820}}.

\bibitem{pythia}
T.~Sjostrand, S.~Mrenna, and P.~Z. Skands, {\em {PYTHIA 6.4 physics and
  manual}}, \href{http://dx.doi.org/10.1088/1126-6708/2006/05/026}{JHEP
  {\bfseries 0605} (2006) 026}.

\bibitem{ct10}
{H.~L.~Lai et al.}, {\em {New parton distributions for collider physics}},
  \href{http://dx.doi.org/10.1103/PhysRevD.82.074024}{Phys. Rev. {\bfseries D
  82} (2010) 074024}, \href{http://arxiv.org/abs/1007.2241}{{\ttfamily
  arXiv:1007.2241}}.

\bibitem{au2}
{ATLAS Collaboration}, {\em Summary of ATLAS Pythia 8 tunes},
  \href{https://cds.cern.ch/record/1474107}{ATL-PHYS-PUB-2012-003} (2012).

\bibitem{deFlorian:2012mx}
D.~de~Florian, G.~Ferrera, M.~Grazzini, and D.~Tommasini, {\em {Higgs boson
  production at the LHC: transverse momentum resummation effects in the
  H$\to$2gamma, H$\to$WW$\to$lnu lnu and H$\to$ZZ$\to$4l decay modes}},
  \href{http://dx.doi.org/10.1007/JHEP06(2012)132}{JHEP {\bfseries 1206} (2012)
  132},
\href{http://arxiv.org/abs/1203.6321}{{\ttfamily arXiv:1203.6321}}.

\bibitem{Grazzini:2013mca}
M.~Grazzini and H.~Sargsyan, {\em {Heavy-quark mass effects in Higgs boson
  production at the LHC}},
  \href{http://dx.doi.org/10.1007/JHEP09(2013)129}{JHEP {\bfseries 1309} (2013)
  129},
\href{http://arxiv.org/abs/1306.4581}{{\ttfamily arXiv:1306.4581}}.

\bibitem{cteq6}
{P.~M.~Nadolsky et al.}, {\em {Implications of CTEQ global analysis for
  collider observables}},
  \href{http://dx.doi.org/10.1103/PhysRevD.78.013004}{Phys. Rev. {\bfseries D
  78} (2008) 013004}, \href{http://arxiv.org/abs/0802.0007}{{\ttfamily
  arXiv:0802.0007}}.

\bibitem{Heinemeyer:2013tqa}
{LHC Higgs Cross Section Working Group}, S.~Dittmaier, C.~Mariotti,
  G.~Passarino, and R.~Tanaka~(Eds.), {\em {Handbook of LHC Higgs Cross
  Sections: 3. Higgs Properties}},
\href{http://arxiv.org/abs/1307.1347}{{\ttfamily arXiv:1307.1347}}.

\bibitem{Djouadi:1991tka}
A.~Djouadi, M.~Spira, and P.~Zerwas, {\em {Production of Higgs bosons in proton
  colliders: QCD corrections}},
\href{http://dx.doi.org/10.1016/0370-2693(91)90375-Z}{Phys. Lett. {\bfseries B
  264} (1991) 440--446}.

\bibitem{Dawson:1990zj}
S.~Dawson, {\em {Radiative corrections to Higgs boson production}},
\href{http://dx.doi.org/10.1016/0550-3213(91)90061-2}{Nucl. Phys. {\bfseries B
  359} (1991) 283--300}.

\bibitem{Spira:1995rr}
M.~Spira, A.~Djouadi, D.~Graudenz, and P.~Zerwas, {\em {Higgs boson production
  at the LHC}}, \href{http://dx.doi.org/10.1016/0550-3213(95)00379-7}{Nucl.
  Phys. {\bfseries B 453} (1995) 17--82},
\href{http://arxiv.org/abs/hep-ph/9504378}{{\ttfamily arXiv:hep-ph/9504378}}.

\bibitem{Harlander:2002wh}
R.~V. Harlander and W.~B. Kilgore, {\em {Next-to-next-to-leading order Higgs
  production at hadron colliders}},
  \href{http://dx.doi.org/10.1103/PhysRevLett.88.201801}{Phys. Rev. Lett.
  {\bfseries 88} (2002) 201801},
\href{http://arxiv.org/abs/hep-ph/0201206}{{\ttfamily arXiv:hep-ph/0201206}}.

\bibitem{Anastasiou:2002yz}
C.~Anastasiou and K.~Melnikov, {\em {Higgs boson production at hadron colliders
  in NNLO QCD}}, \href{http://dx.doi.org/10.1016/S0550-3213(02)00837-4}{Nucl.
  Phys. {\bfseries B 646} (2002) 220--256},
\href{http://arxiv.org/abs/hep-ph/0207004}{{\ttfamily arXiv:hep-ph/0207004}}.

\bibitem{Ravindran:2003um}
V.~Ravindran, J.~Smith, and W.~L. van Neerven, {\em {NNLO corrections to the
  total cross-section for Higgs boson production in hadron hadron collisions}},
  \href{http://dx.doi.org/10.1016/S0550-3213(03)00457-7}{Nucl. Phys. {\bfseries
  B 665} (2003) 325--366},
\href{http://arxiv.org/abs/hep-ph/0302135}{{\ttfamily arXiv:hep-ph/0302135}}.

\bibitem{Aglietti2004}
U.~Aglietti, R.~Bonciani, G.~Degrassi, and A.~Vicini, {\em {Two-loop light
  fermion contribution to Higgs production and decays}},
  \href{http://dx.doi.org/10.1016/j.physletb.2004.06.063}{Phys. Lett.
  {\bfseries B 595} (2004) 432},
\href{http://arxiv.org/abs/hep-ph/0404071}{{\ttfamily arXiv:hep-ph/0404071}}.

\bibitem{Actis2008}
S.~Actis, G.~Passarino, C.~Sturm, and S.~Uccirati, {\em {NLO Electroweak
  Corrections to Higgs Boson Production at Hadron Colliders}},
  \href{http://dx.doi.org/10.1016/j.physletb.2008.10.018}{Phys. Lett.
  {\bfseries B 670} (2008) 12},
\href{http://arxiv.org/abs/0809.1301}{{\ttfamily arXiv:0809.1301}}.

\bibitem{Ciccolini:2007jr}
M.~Ciccolini, A.~Denner, and S.~Dittmaier, {\em {Strong and electroweak
  corrections to the production of Higgs + 2-jets via weak interactions at the
  LHC}}, \href{http://dx.doi.org/10.1103/PhysRevLett.99.161803}{Phys. Rev.
  Lett. {\bfseries 99} (2007) 161803},
\href{http://arxiv.org/abs/0707.0381}{{\ttfamily arXiv:0707.0381}}.

\bibitem{Ciccolini:2007ec}
M.~Ciccolini, A.~Denner, and S.~Dittmaier, {\em {Electroweak and QCD
  corrections to Higgs production via vector-boson fusion at the LHC}},
  \href{http://dx.doi.org/10.1103/PhysRevD.77.013002}{Phys. Rev. {\bfseries D
  77} (2008) 013002},
\href{http://arxiv.org/abs/0710.4749}{{\ttfamily arXiv:0710.4749}}.

\bibitem{Arnold:2008rz}
K.~Arnold {et~al.}, {\em {VBFNLO: A Parton level Monte Carlo for processes with
  electroweak bosons}},
  \href{http://dx.doi.org/10.1016/j.cpc.2009.03.006}{Comput.Phys.Commun.
  {\bfseries 180} (2009) 1661--1670},
\href{http://arxiv.org/abs/0811.4559}{{\ttfamily arXiv:0811.4559}}.

\bibitem{Bolzoni:2010xr}
P.~Bolzoni, F.~Maltoni, S.-O. Moch, and M.~Zaro, {\em {Higgs production via
  vector-boson fusion at NNLO in QCD}},
  \href{http://dx.doi.org/10.1103/PhysRevLett.105.011801}{Phys. Rev. Lett.
  {\bfseries 105} (2010) 011801},
\href{http://arxiv.org/abs/1003.4451}{{\ttfamily arXiv:1003.4451}}.

\bibitem{Han:1991ia}
T.~Han and S.~Willenbrock, {\em {QCD correction to the $pp \rightarrow WH$ and
  $ZH$ total cross-sections}},
\href{http://dx.doi.org/10.1016/0370-2693(91)90572-8}{Phys. Lett. {\bfseries B
  273} (1991) 167--172}.

\bibitem{Brein:2003wg}
O.~Brein, A.~Djouadi, and R.~Harlander, {\em {NNLO QCD corrections to the
  Higgs-strahlung processes at hadron colliders}},
  \href{http://dx.doi.org/10.1016/j.physletb.2003.10.112}{Phys. Lett.
  {\bfseries B 579} (2004) 149--156},
\href{http://arxiv.org/abs/hep-ph/0307206}{{\ttfamily arXiv:hep-ph/0307206}}.

\bibitem{Ciccolini:2003jy}
M.~Ciccolini, S.~Dittmaier, and M.~{Kramer}, {\em {Electroweak radiative
  corrections to associated $WH$ and $ZH$ production at hadron colliders}},
  \href{http://dx.doi.org/10.1103/PhysRevD.68.073003}{Phys. Rev. {\bfseries D
  68} (2003) 073003},
\href{http://arxiv.org/abs/hep-ph/0306234}{{\ttfamily arXiv:hep-ph/0306234}}.

\bibitem{Djouadi:1997yw}
A.~Djouadi, J.~Kalinowski, and M.~Spira, {\em {HDECAY: A Program for Higgs
  boson decays in the standard model and its supersymmetric extension}},
  \href{http://dx.doi.org/10.1016/S0010-4655(97)00123-9}{Comput. Phys. Commun.
  {\bfseries 108} (1998) 56--74},
\href{http://arxiv.org/abs/hep-ph/9704448}{{\ttfamily arXiv:hep-ph/9704448}}.

\bibitem{Mangano:2002ea}
M.~L. Mangano, M.~Moretti, F.~Piccinini, R.~Pittau, and A.~D. Polosa, {\em
  {ALPGEN, a generator for hard multiparton processes in hadronic collisions}},
  JHEP {\bfseries 0307} (2003) 001,
\href{http://arxiv.org/abs/hep-ph/0206293}{{\ttfamily arXiv:hep-ph/0206293}}.

\bibitem{Corcella:2000bw}
G.~Corcella {et~al.}, {\em {HERWIG 6: An Event generator for hadron emission
  reactions with interfering gluons (including supersymmetric processes)}},
  JHEP {\bfseries 0101} (2001) 010,
\href{http://arxiv.org/abs/hep-ph/0011363}{{\ttfamily arXiv:hep-ph/0011363}}.

\bibitem{mcatnlo}
S.~Frixione, P.~Nason, and B.~R. Webber, {\em {Matching NLO QCD and parton
  showers in heavy flavor production}}, JHEP {\bfseries 0308} (2003) 007,
\href{http://arxiv.org/abs/hep-ph/0305252}{{\ttfamily arXiv:hep-ph/0305252}}.

\bibitem{Kersevan:2004yg}
B.~P. Kersevan and E.~Richter-Was, {\em {The Monte Carlo event generator AcerMC
  version 2.0 with interfaces to PYTHIA 6.2 and HERWIG 6.5}},
\href{http://arxiv.org/abs/hep-ph/0405247}{{\ttfamily arXiv:hep-ph/0405247}}.

\bibitem{gg2WW}
T.~Binoth, M.~Ciccolini, N.~Kauer, and M.~Kramer, {\em {Gluon-induced W-boson
  pair production at the LHC}},
  \href{http://dx.doi.org/10.1088/1126-6708/2006/12/046}{JHEP {\bfseries 0612}
  (2006) 046},
\href{http://arxiv.org/abs/hep-ph/0611170}{{\ttfamily arXiv:hep-ph/0611170}}.

\bibitem{Melia:2011tj}
T.~Melia, P.~Nason, R.~Rontsch, and G.~Zanderighi, {\em {$W^+W^-$, $WZ$ and
  $ZZ$ production in the POWHEG BOX}},
  \href{http://dx.doi.org/10.1007/JHEP11(2011)078}{JHEP {\bfseries 1111} (2011)
  078},
\href{http://arxiv.org/abs/1107.5051}{{\ttfamily arXiv:1107.5051}}.

\bibitem{Alwall:2007st}
{J.~Alwall et al.}, {\em {MadGraph/MadEvent v4: The New Web Generation}}, JHEP
  {\bfseries 0709} (2007) 028, \href{http://arxiv.org/abs/0706.2334}{{\ttfamily
  arXiv:0706.2334}}.

\bibitem{Alwall:2011uj}
J.~Alwall, M.~Herquet, F.~Maltoni, O.~Mattelaer, and T.~Stelzer, {\em {MadGraph
  5 : Going Beyond}}, \href{http://dx.doi.org/10.1007/JHEP06(2011)128}{JHEP
  {\bfseries 06} (2011) 128},
\href{http://arxiv.org/abs/1106.0522}{{\ttfamily arXiv:1106.0522}}.

\bibitem{atlassim}
{ATLAS Collaboration}, {\em {The ATLAS simulation infrastructure}},
  \href{http://dx.doi.org/10.1140/epjc/s10052-010-1429-9}{Eur. Phys. J.
  {\bfseries C 70} (2010) 823},
\href{http://arxiv.org/abs/1005.4568}{{\ttfamily arXiv:1005.4568}}.

\bibitem{geant4}
{GEANT4} Collaboration, S.~Agostinelli {et~al.}, {\em {GEANT4: A Simulation
  toolkit}},
\href{http://dx.doi.org/10.1016/S0168-9002(03)01368-8}{Nucl. Instrum. Meth.
  {\bfseries A 506} (2003) 250--303}.

\bibitem{Cacciari:2008gp}
M.~Cacciari, G.~P. Salam, and G.~Soyez, {\em {The anti-$\text{k}_{\text{t}}$
  jet clustering algorithm}},
  \href{http://dx.doi.org/10.1088/1126-6708/2008/04/063}{JHEP {\bfseries 0804}
  (2008) 063},
\href{http://arxiv.org/abs/0802.1189}{{\ttfamily arXiv:0802.1189}}.

\bibitem{Aad:2014bia}
{ATLAS} Collaboration, {\em {Jet energy measurement and its systematic
  uncertainty in proton--proton collisions at $\sqrt{s}=7$~TeV with the ATLAS
  detector}},
\href{http://arxiv.org/abs/1406.0076}{{\ttfamily arXiv:1406.0076}}.

\bibitem{btag7tev}
{ATLAS} Collaboration, {\em {Commissioning of the ATLAS high-performance
  b-tagging algorithms in the 7 TeV collision data}}, ATLAS-CONF-2011-102,
  \href{http://arxiv.org/abs/http://cdsweb.cern.ch/record/1369219}{{\ttfamily
  http://cdsweb.cern.ch/record/1369219}}.

\bibitem{bperf1}
{ATLAS} Collaboration, {\em {Measuring the $b$-tag efficiency in a $t\bar{t}$
  sample with 4.7~fb$^{-1}$ of data from the ATLAS detector}},
  ATLAS-CONF-2012-097,
  \href{http://arxiv.org/abs/http://cds.cern.ch/record/1460443}{{\ttfamily
  http://cds.cern.ch/record/1460443}}.

\bibitem{Vichou:2012roa}
{ATLAS} Collaboration, {\em {Performance of missing transverse momentum
  reconstruction in proton-proton collisions at $\sqrt{s}$ = 7~TeV with
  ATLAS}},
\href{http://dx.doi.org/10.1088/1742-6596/404/1/012011}{J. Phys. Conf. Ser.
  {\bfseries 404} (2012) 012011}.

\bibitem{Aad:2013ucp}
{ATLAS} Collaboration, {\em {Improved luminosity determination in $pp$
  collisions at $\sqrt{s}$ = 7 TeV using the ATLAS detector at the LHC}},
  \href{http://dx.doi.org/10.1140/epjc/s10052-013-2518-3}{Eur.Phys.J.
  {\bfseries C 73} (2013) 2518},
\href{http://arxiv.org/abs/1302.4393}{{\ttfamily arXiv:1302.4393}}.

\bibitem{Cowan:2010js}
G.~Cowan, K.~Cranmer, E.~Gross, and O.~Vitells, {\em {Asymptotic formulae for
  likelihood-based tests of new physics}},
  \href{http://dx.doi.org/10.1140/epjc/s10052-011-1554-0}{Eur. Phys. J.
  {\bfseries C 71} (2011) 1554},
\href{http://arxiv.org/abs/1007.1727}{{\ttfamily arXiv:1007.1727}}.

\bibitem{Read:2002hq}
A.~L. Read, {\em {Presentation of search results: The CL(s) technique}},
\href{http://dx.doi.org/10.1088/0954-3899/28/10/313}{J. Phys. {\bfseries G 28}
  (2002) 2693--2704}.

\end{thebibliography}\endgroup

\clearpage
\onecolumn
\begin{flushleft}
{\Large The ATLAS Collaboration}

\bigskip

G.~Aad$^{\rm 84}$,
B.~Abbott$^{\rm 112}$,
J.~Abdallah$^{\rm 152}$,
S.~Abdel~Khalek$^{\rm 116}$,
O.~Abdinov$^{\rm 11}$,
R.~Aben$^{\rm 106}$,
B.~Abi$^{\rm 113}$,
M.~Abolins$^{\rm 89}$,
O.S.~AbouZeid$^{\rm 159}$,
H.~Abramowicz$^{\rm 154}$
% ...
\bigskip
\\
% ...
$^{11}$ Institute of Physics, Azerbaijan Academy of Sciences, Baku, Azerbaijan\\
% ...
$^{84}$ CPPM, Aix-Marseille Universit{\'e} and CNRS/IN2P3, Marseille, France\\
% ...
$^{89}$ Department of Physics and Astronomy, Michigan State University, East Lansing MI, United States of America\\
% ...
$^{106}$ Nikhef National Institute for Subatomic Physics and University of Amsterdam, Amsterdam, Netherlands\\
% ...
$^{112}$ Homer L. Dodge Department of Physics and Astronomy, University of Oklahoma, Norman OK, United States of America\\
$^{113}$ Department of Physics, Oklahoma State University, Stillwater OK, United States of America\\
% ...
$^{116}$ LAL, Universit{\'e} Paris-Sud and CNRS/IN2P3, Orsay, France\\
% ...
$^{152}$ Institute of Physics, Academia Sinica, Taipei, Taiwan\\
% ...
$^{154}$ Raymond and Beverly Sackler School of Physics and Astronomy, Tel Aviv University, Tel Aviv, Israel\\
% ...
$^{159}$ Department of Physics, University of Toronto, Toronto ON, Canada\\
% ...
\end{flushleft}


\end{document}